# MANIFESTATION OF DEFECTS IN RAMAN EFFECT OF LIGHT SPECTRA OF ORGANIC MOLECULAR CRYSTALS (COMPUTER MODELING)

M.A.Korshunov[*]

*L.V. Kirensky Institute of physics, Siberian branch of the Russian Academy of Sciences, 660036 Krasnoyarsk, Russia*

**ABSTRACT.** Calculations of frequency spectra for organic molecular crystals taking various defects in their structure into account are carried out. It is shown how the presence of the vacancies, orientation disorder, and surface presence affects spectra. It was found, that each defect has the characteristic set of lines in the spectra.

At examination Raman of spectra of organic molecular crystals is observed occurrence of some lines forbidden by the selection rules. These lines often have small intensity and demand additional examinations, what them to reveal. But these lines appear constantly for various samples of one substance and on different equipment, and to carry them to display of noise inconveniently. For example, in para-dichlorobenzol [1] there are additional lines of small intensity in a spectrum of the lattice oscillations gained from different samples and on different devices. Thus, values of frequencies of these lines coincide. However, they do not coincide with values of frequencies of orientation and transmitting lines. It is supposed that violation of selection rules is caused by presence of defects [2]. However, how the presence of those or other defects affects spectra demands additional studying. In present work, the results of calculations of spectra of small frequencies of para-dichlorobenzol are given at presence in structure of vacancies, orientation disorder and surface presence. On X-ray diffraction data the monocrystal α - para-dichlorobenzol phases crystallises in space group P2$_1$/a with two molecules in an unit cell [3]. In a spectrum of the lattice oscillations of such ideal crystals it should be observed six lines, the molecules caused by circumrotatory oscillations. At calculations interaction between molecules paid off on a method atom-atom of potentials. Interaction coefficients got out such that the spectrum of the lattice oscillations of an ideal crystal coincided with the observational spectrum of small frequencies. Further at calculations of crystals with defects they did not vary.

In Fig.1 the observational spectrum of polycrystalline para-dichlorobenzol (a) and the calculated spectrum (b) is shown. Frequencies of transmitting oscillations are figured by lines of smaller quantity.

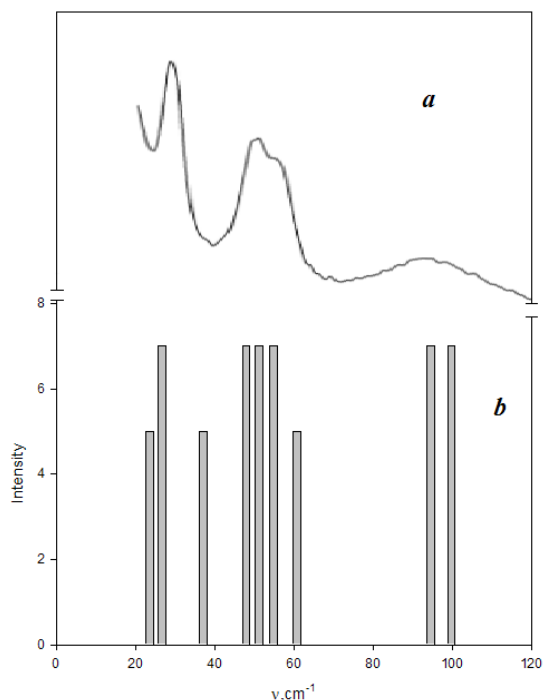

**Fig. 1.** The observational spectrum of para-dichlorobenzol (a) and the calculated spectrum (b) an ideal crystal.

The first defect inlet into structure is vacancy. Calculations were spent with use of a method the Dyne [4]. This method allows to carry out calculations of frequency spectra of disorder molecular crystals. On the basis of calculations histograms which show probability of display of lines of a spectrum in the chosen frequency interval have been gained.

In Fig. 2 the histogram of a spectrum of para-dichlorobenzol is shown at the account in structure of vacancies. As we see unlike an ideal crystal in a spectrum there are additional lines both below 20 cm$^{-1}$, and in the field of 80 cm$^{-1}$.

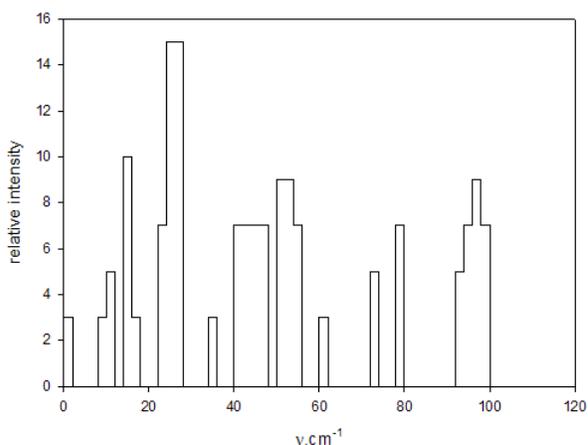

Fig. 2. The histogram of a spectrum of para-dichlorobenzol at presence in structure of vacancies.

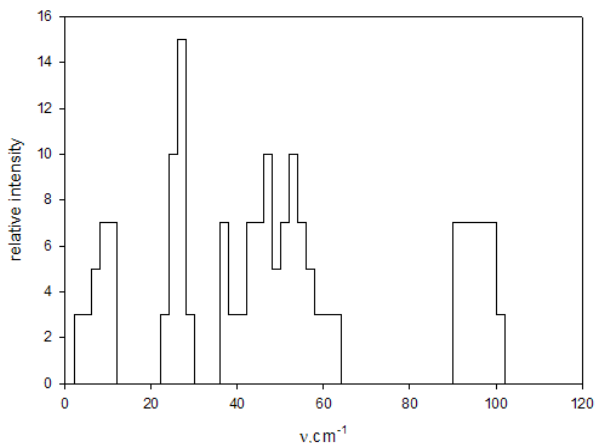

Fig. 3. The histogram of a spectrum of para-dichlorobenzol at the account of orientation disorder of molecules.

Orientation disorder of molecules that can be caused also presence of vacancies or defects of packing was the second viewed defect.

In Fig. 3 the histogram taking into account orientation disorder of molecules from $0^0$ to $3^0$ is shown.

As we see additional lines have moved more close to raising and the spectrum line broadening is observed.

The third defect viewed in article, surface presence that it is necessary to consider at interpretation of spectra is. As spectra are gained from crystals of the terminating size in which the surface can play an essential role at reduction of the sizes of a crystal. In Fig. 4 effects of calculations taking into account presence of a surface and the account of the superficial oscillations are shown. As we see in a spectrum also there are additional lines, but below 15 cm$^{-1}$ and in the field of 80 cm$^{-1}$ they miss.

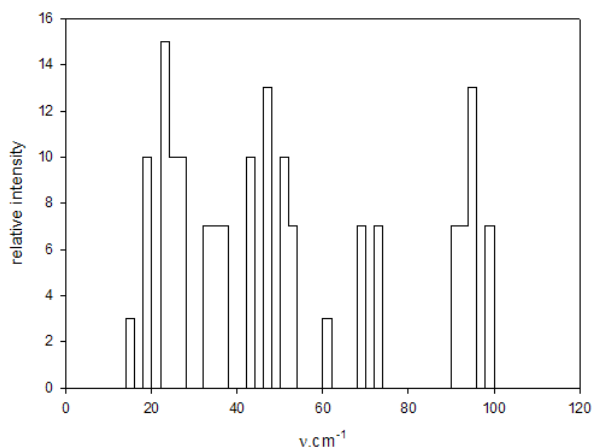

Fig. 4. The histogram of a spectrum of para-dichlorobenzol at the surface account as defect.

Thus, calculations have shown, that each defect has the display in a spectrum of small frequencies, it allows to interpret a real spectrum and to speak about presence of those or other defects.